\makeatletter
\declare@file@substitution{revtex4-1.cls}{revtex4-2.cls}
\makeatother

\documentclass[twocolumn]{aastex631}

\usepackage{longtable}
\usepackage{afterpage}

\shorttitle{V965 Cephei revisited: evidence for a binary system}
\shortauthors{Pyatnytskyy}
\graphicspath{{./}{figures/}}
\begin{document}

\title{V965 Cephei revisited: evidence for a binary system}

\correspondingauthor{Maksym Pyatnytskyy}
\email{pmak@osokorky-observatory.com}

\author[0000-0002-0813-1000]{Maksym Pyatnytskyy}
\affiliation{Private Observatory ''Osokorky'', P. O. Box 27, Kyiv, 02132, Ukraine}

\begin{abstract}
We extended the $O-C$ diagram of the HADS star V965 Cep with all currently available observations in the Johnson V filter and added unfiltered ones.
The new, up-to-date $O-C$ diagram shows that the seeming period change previously revealed by the author does not occur uniformly.
Instead, the near-parabolic part of the $O-C$ diagram can be a part of a periodic curve.
This could be a sign of a second body in the system.
Based on our observations, we also estimated the $B-V$ index of the star to be $ 0.48 \pm 0.05$ at maximum brightness and $ 0.57 \pm 0.05$ at minimum brightness.
\end{abstract}


\section{Introduction}

The star V965 Cep was discovered during the NSVS survey \citep{2004AJ....127.2436W}; the proper classification of it as a High Amplitude Delta Scuti (HADS) star was done by \citet{2009PZP.....9...30S}.
The author already investigated the star with colleagues \citep{2021JAVSO..49...58P, 2024RNAAS...8..159P}.
We examined how the differences between the observed and calculated ($O-C$) times of brightness maxima depend on the cycle number. The graphical representation of this relationship is commonly referred to as the $O-C$ diagram, which helps reveal deviations from the model that assumes a constant oscillation frequency. A comprehensive description of the approach can be found in \citet{2005ASPC..335....3S}.

It was found \citep{2021JAVSO..49...58P} that an $O-C$ diagram for the moments of the star's maximum light can be quite satisfactorily approximated with a parabolic curve in the interval from September 2011 to December 2020.
To build the mentioned $O-C$ diagram, the author used observations in the Johnson V filter from the AAVSO International Database (AID) \url{http://www.aavso.org}, from the literature \citep{2012IBVS.6015....1W, 2015IBVS.6150....1W}, and his own observations (currently reported to the AID, AAVSO observer code PMAK).
The restriction to only the Johnson V filter was done intentionally (even though there were many observations in the 'clear' filter (i.e., without any filter at all)): there were concerns that the form of the light curve (and, therefore, the moments of maxima) may differ in the different filters.
In recent years, however, most of the observations available in the AID have been reported in the CV band (unfiltered with V zeropoint), i.e., without any photometric filter.
Even his own latest observations the author conducted without filter to increase the signal-to-noise ratio.
In the current work, the author successfully utilized all available observations, both with and without filters, to extend the $O-C$ diagram.
This led to the complete reinterpretation of it.

\section{Methods}

\subsection{The Author's Observations}

The author observed V965 Cep with a 6-inch f/5 Newtonian equipped with a cooled CMOS camera ASI183MM Pro.
Most observations were conducted without a photometric filter; however, simultaneous observations in the Johnson V and B filters were carried out during three nights in January, August, and September 2023.
One maximum was also observed in August 2021 with the Johnson V filter only.
The observational site was in the Osokorky neighborhood, on the left bank of the Dnieper River, in Kyiv, Ukraine. According to the light pollution map, the Bortle class for that place is 6, corresponding to the "bright suburban" sky.
The author used his own command-line toolkit to calibrate the images: \url{https://github.com/mpyat2/MaxFITStoolkit}.
The differential aperture photometry was performed with AstroImageJ \citep{2017AJ....153...77C}.
A comparison and check star were the same as in \citet{2021JAVSO..49...58P}.

\subsection{Deriving Times of Maxima}
To derive times of maxima (ToMs) from the observed light curves, the author used an approach described by \citet{2009IBVS.5878....1W}: fitting a model light curve to the observed one.
This approach is based on the assumption that the shape of maxima does not change from cycle to cycle.
Although this is not precisely true for V965 Cep (see \citep{2024RNAAS...8..159P}), the changes in the shape are pretty subtle and can be neglected.

Since the amplitudes and shapes of the maxima differ in different bands (see below the light curves in the V and B filters), a separate model was used for each photometric filter.
The current research used the same model as in \citet{2021JAVSO..49...58P} to analyze the light curves in the V filter.
For the "clear" filter, the model was built using only the author's observations (carried out during 2023 and 2024).
This was done to minimize extra uncertainties caused by somewhat different zero points for different observers and possible other observer-dependent peculiarities.

The model light curves were built using the VStar software \citep{2012JAVSO..40..852B, Benn_VStar_2024} with the "Fourier Model" tool.
This tool approximates the light curve with a set of trigonometric polynomials. We used a trigonometric polynomial composed of the sinusoidal wave with the primary frequency $f_0$ on which the star oscillates, which is equal to $1/P$, where $P$ is a period of variability, and its harmonics up to $s \cdot f_0$ ($s$ is a positive integer greater than 1).

The question arose: what was the maximum number of harmonics that should be used for the approximation?
To answer this question, an approach described in \citet{1994OAP.....7...49A, 2020kdbd.book..191A} was used to determine the statistically optimal approximation, which corresponds to the minimal r.m.s. accuracy of the approximations.
An estimation of such an optimal approximation was based on the following formula:
\begin{equation}
\sigma_m^2[x_C]=\frac{m}{n(n-m)}\sum_{k=1}^n (x_k - x_{Ck})^2
\end{equation}
where $n$ is the number of observations, $m$ is the number of parameters. In our case, it is equal to $2s+1$, $x_k$ is a $k$-th observation, $x_{Ck}$ is a $k$-th calculated (model) point.

\begin{figure}[ht!]
\begin{center}
\includegraphics[scale=0.5,angle=0]{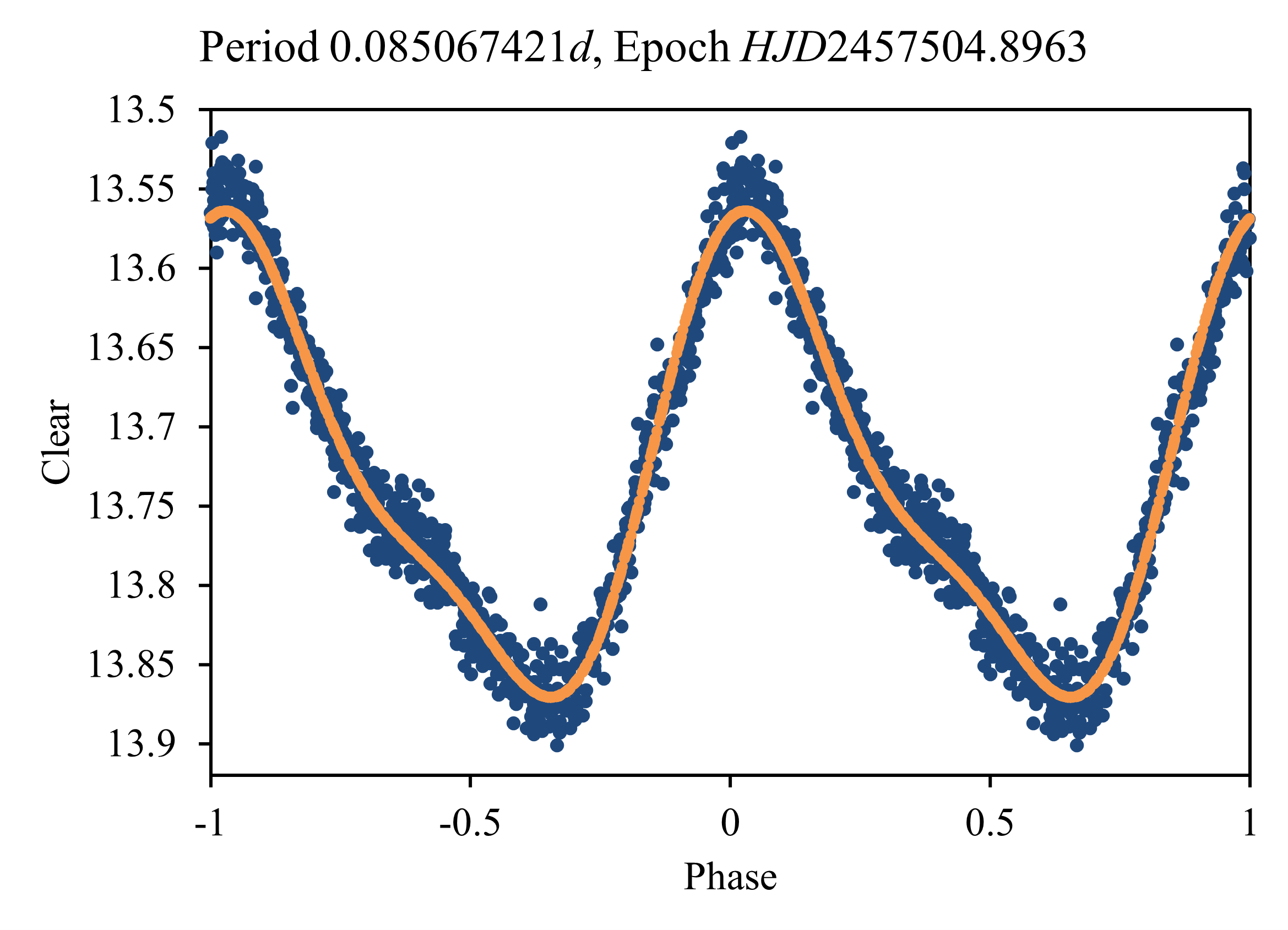}
\caption{
The phase diagram for the observations in the "clear" filter conducted by the author during 2023 and up to June 18, 2024. The Fourier model is shown with the light-color line.
\label{fig:1}}
\end{center}
\end{figure}

Using this approach, it was defined that the second harmonic ($2 \cdot f_0$) only is needed, besides the main frequency $f_0$, to adequately approximate the observations: the $\sigma_m^2[x_C]$ value minimizes for $s=2$. 
The approximation is shown in Fig.~\ref{fig:1}. This model was used to find maxima in the observations in the "clear" filter.
\pagebreak
\section{Results and Discussion}

\subsection{The BV photometry}

\begin{figure}[ht!]
\begin{center}
\includegraphics[scale=0.5,angle=0]{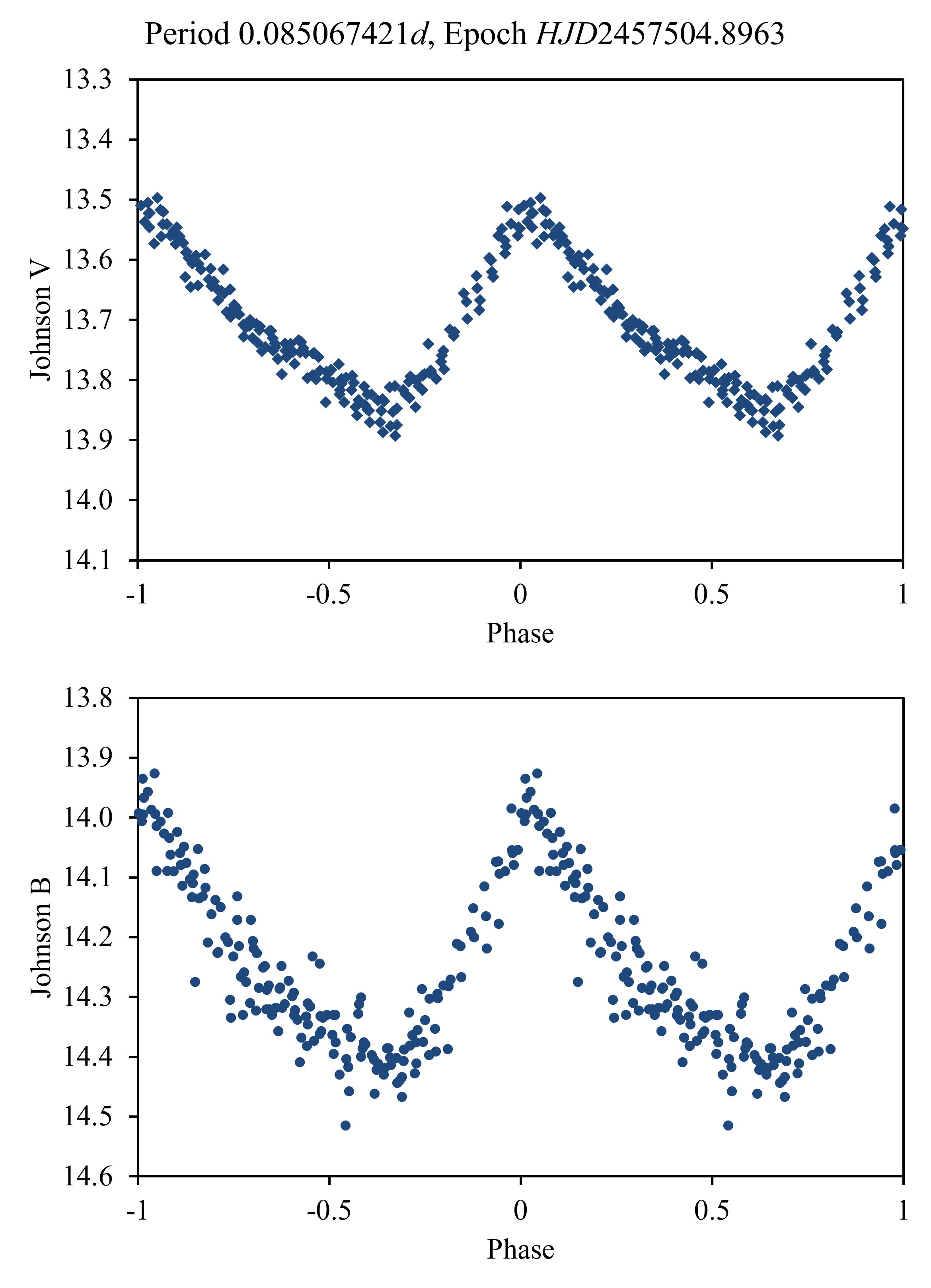}
\caption{
The phase diagrams for the author's observations in the V and B bands.
\label{fig:2}}
\end{center}
\end{figure}

As it was previously indicated, over three nights, V and B filters were concurrently used. Phased light curves for these observations are shown in Fig.~\ref{fig:2}.

\begin{figure}[ht!]
\begin{center}
\includegraphics[scale=0.7,angle=0]{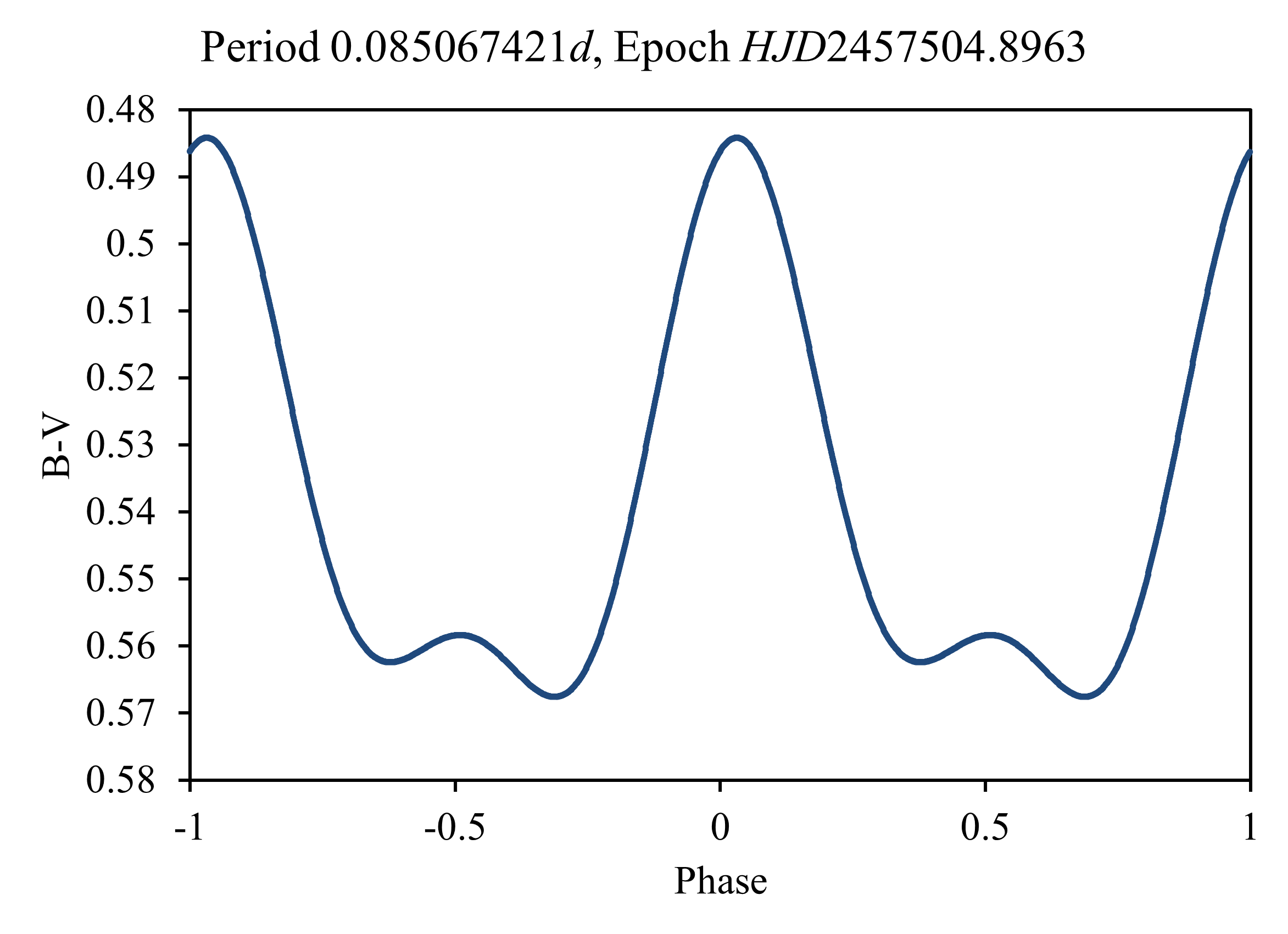}
\caption{
The phase diagram for the B-V model curve.
\label{fig:3}}
\end{center}
\end{figure}

The scatter is pretty large, especially for the data in the B filter, so the direct subtraction gives a too noisy $B-V$ light curve. So, Fourier models were created for these curves, as described above.
The model $B-V$ curve shown in Fig.~\ref{fig:3} was calculated from these two models.
Our estimation for the $B-V$ index is $ 0.48 \pm 0.05$ at the maximum and $ 0.57 \pm 0.05$ at the minimum, respectively. 
These values differ noticeably from the $B-V$ index value listed in the International Variable Star Index (VSX) \citep{2014yCat....102027W}.
The VSX value ($0.44$) is derived from the APASS DR9 survey \citep{2015AAS...22533616H}, which has a notable uncertainty of $0.16$. Our values fall within the uncertainty range reported by the VSX.

Variations in the color index over the pulsation cycle reflect changes in effective temperature. These variations may influence the measurement accuracy, particularly if the comparison star has a substantially different color index and the transformation coefficient of the observational setup (camera and filter) differs significantly from zero. Fortunately, the transformation coefficients for both the V and B filters are small in our case. For the V filter, the transformation coefficient $T_{V(B-V)}$ ranges from $0.009$ (based on observations of the M67 AAVSO Standard Field; \citealt{2021JAVSO..49...58P}) to $-0.02$ (SA32 standard field). For the B filter, the transformation coefficient derived from the observations of the SA32 standard field is $-0.02$. The $B-V$ color index of the comparison star is $0.95$ (see Table~1 in \cite{2021JAVSO..49...58P}). Thus, for both the V and B filters, the systematic bias caused by the color difference between the comparison and target stars is less than or approximately 0.01 magnitude in absolute value, which is significantly smaller than the random error and therefore can be neglected.

\subsection{The $O-C$ Diagram}
For building the $O-C$ diagram, times of maxima (ToMs) in the V filter listed in  \citep{2021JAVSO..49...58P} were used along with several new ToMs in the V filter from the recent observations, ToMs in the "clear" filter from the literature \citep{2011IBVS.5977....1W, 2012IBVS.6015....1W, 2013IBVS.6049....1W} and from the AID, including the author's ones.
New ToMs not included in the previous paper \citep{2021JAVSO..49...58P}, are listed in Table~\ref{tab1}.

\begin{figure}[ht!]
\begin{center}
\includegraphics[scale=0.7,angle=0]{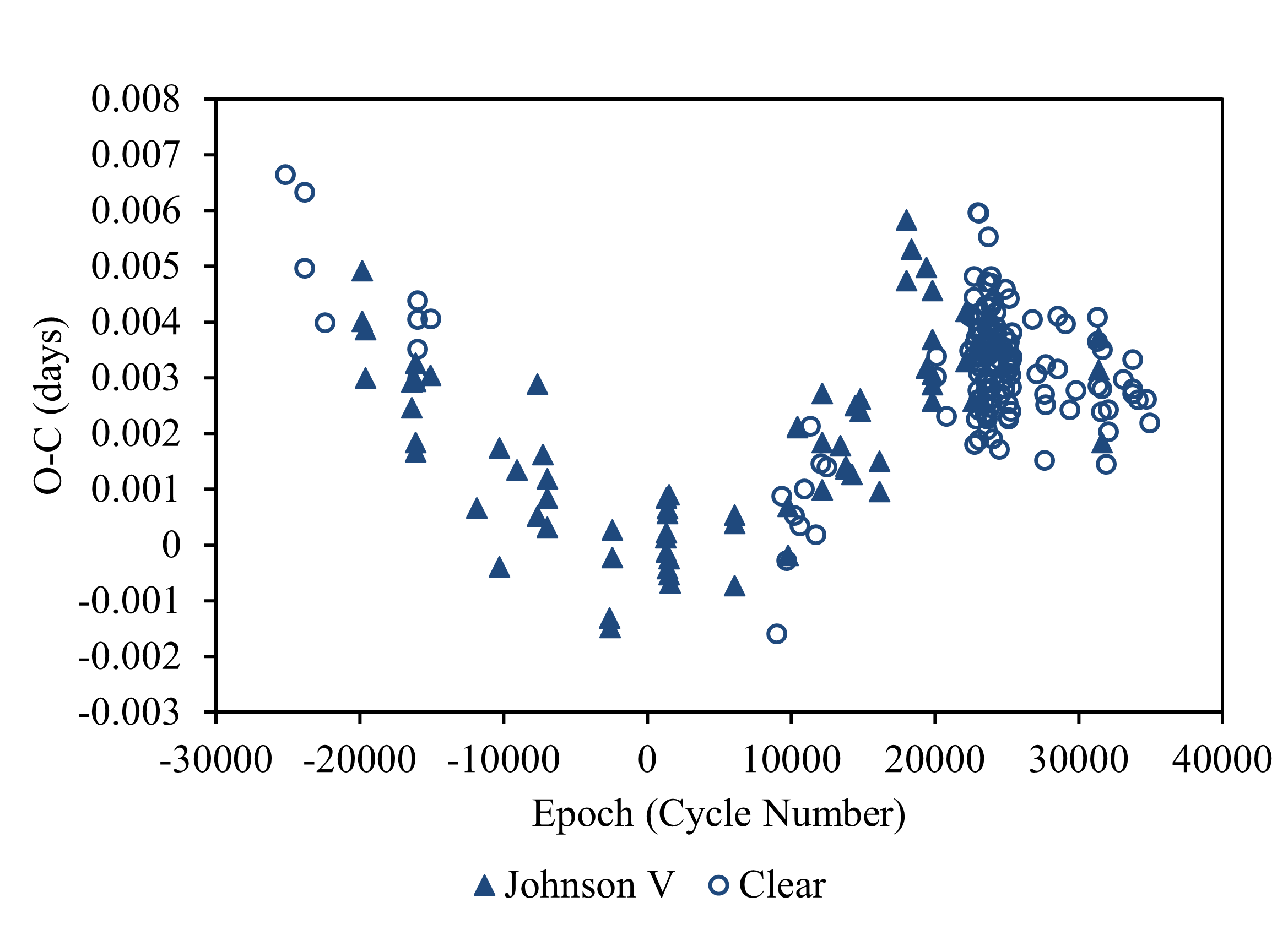}
\caption{
The O-C diagram for V965 Cep.
\label{fig:4}}
\end{center}
\end{figure}

The $O-C$ diagram built using the ToMs from \citet{2021JAVSO..49...58P} and from Table~\ref{tab1} is shown in Fig.~\ref{fig:4}. The $O-C$ values were calculated using the period $0.085067421d$ and the initial epoch $HJD2457504.8963$ from \citet{2021JAVSO..49...58P}.

It can be seen that the two sets of data -- in the Johnson V filter and in the clear filter -- are in fairly good agreement.
Noteworthy is the rather large scatter of points.
Although the scatter mainly results from considerable uncertainty in measuring ToMs, the intrinsic causes can partially explain it.
In our work \citep{2024RNAAS...8..159P}, we discovered a subtle modulation of both the period and the amplitude of the light curve; the modulation of the period contributes to the scatter.

It is obvious that the shape of the $O-C$ diagram cannot be explained by a monotonous change in the period.
One may assume that the observed part of the $O-C$ diagram is a part of a periodic curve.

\begin{figure}[ht!]
\begin{center}
\includegraphics[scale=0.7,angle=0]{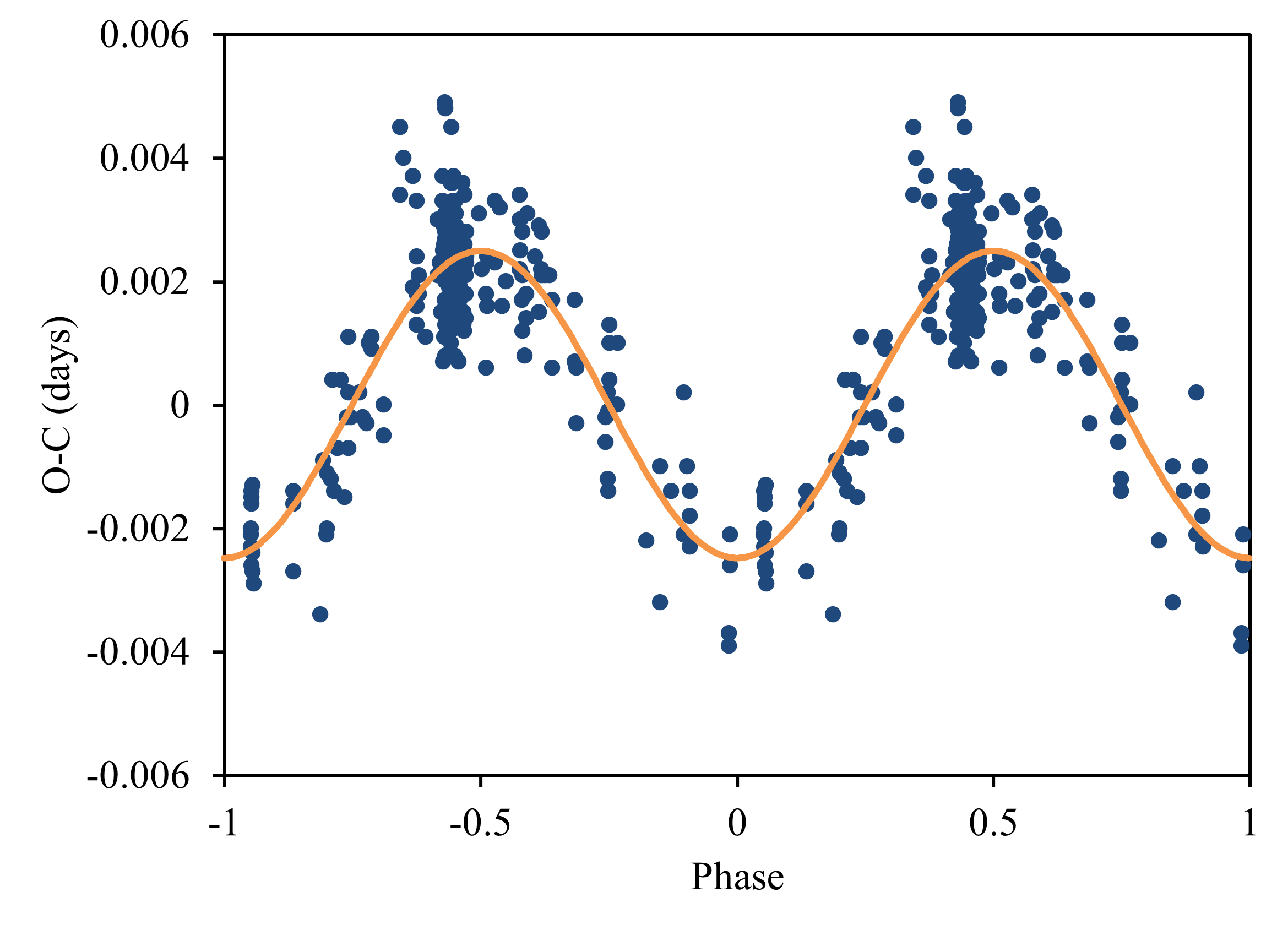}
\caption{
Folded $O-C$ (period $13.4 yr$) diagram after removing a linear trend. The smooth curve is a sinusoidal model.
\label{fig:5}}
\end{center}
\end{figure}

Using the MCV software \citep{2004AstSR...5..264A, 2020kdbd.book..191A}, a period of this hypothetical curve was estimated; the result is shown in Fig.~\ref{fig:5} as a phase diagram. A periodic component of the estimation has a period $\Pi$ of $13.4 \pm 0.9 yr$ and a semi-amplitude of $0.00249 \pm 0.00014 d$.

Such a periodic pattern in the $O-C$ diagram can indicate the presence of a second body in the system.
If this is the case, the observed apparent change in the period is due to a light-time effect in the binary system \citep{2005ASPC..335...85B, 2005JAVSO..34....1T}. By multiplying the semi-amplitude of the model $O-C$ curve ($0.00249 d$) by the speed of light, we find that a hypothetical second body causes a displacement of the pulsating star along the line of sight by ${a = 0.43}$ A.U. in both directions.

Let's assume the orbit is circular.
If so, from the Kepler's equation
\begin{equation}
\sin^3i (M + m) \Pi^2  = (a + A)^3
\end{equation}
where $i$ is the orbit inclination, $M$ is the mass of the observed star, $m$ is the mass of a supposed second body, $A = a M/m$, we can estimate the lower limit of a second body mass if we know the mass of the observed star.

Assuming a pulsating star mass of $1.8M_{\odot}$ (see \citet{2022yCat.1355....0G}), we estimate the lower limit of a second body's mass to be $\approx0.1M_{\odot}$. Thus, a second body could be a faint red dwarf. The lower limit of the semimajor axis $(a + A) \approx7$ A.U.

\section{Conclusion}

The $O-C$ diagram for the V965 Cep demonstrates peculiarities that can be caused by a second body orbiting the common center of mass.
The presence of such a companion causes the observed star to be displaced along the line of sight with a period of $13.4 \pm 0.9 yr$ by a distance of ${a = 0.43 \pm 0.02}$ A.U. in both directions.
Assuming the orbits are circular and the mass of the observed star is about $1.8M_{\odot}$, we estimated a second body mass as $\approx0.1M_{\odot}$.

We also estimated the $B-V$ index for the V965 Cep in the maximum and the minimum of the star's brightness: $ 0.48 \pm 0.05$ and $ 0.57 \pm 0.05$ respectively.

\section*{Acknowledgments}
We acknowledge with thanks the variable star observations from the AAVSO International Database contributed by observers DFS, HMB, SDOA, and VMT, which were used in this research.
The author would like to thank Prof. Ivan L. Andronov for his invaluable advice.

\bibliography{MP_V965_Cep_2024}{}
\bibliographystyle{aasjournal}

\afterpage{
\begin{longtable}[c]{ c  c  c  c  c  c  c }
\caption{Times of maxima for V965 Cep} \\
\hline
Time of Maximum & Uncertainty & Epoch & $O-C$ & Band & Source & AAVSO \\
$(HJD - 2400000)$ & (d) & (Cycle number) & (d) &  &  & Observer Code \\
\hline \hline
59390.5040 & 0.0017 & 22166 & 0.0033 & V & AAVSO & DFS \\
59390.5900 & 0.0017 & 22167 & 0.0042 & V & AAVSO & DFS \\
59428.3583 & 0.0024 & 22611 & 0.0026 & V & AAVSO & PMAK \\
60176.3573 & 0.0011 & 31404 & 0.0037 & V & AAVSO & PMAK \\
60176.4418 & 0.0011 & 31405 & 0.0031 & V & AAVSO & PMAK \\
60195.3255 & 0.0023 & 31627 & 0.0018 & V & AAVSO & PMAK \\
55365.4573 & 0.0013 & -25150 & 0.0066 & C & \citep{2011IBVS.5977....1W} & - \\
55479.2772 & 0.0016 & -23812 & 0.0063 & C & \citep{2011IBVS.5977....1W} & - \\
55479.3609 & 0.0009 & -23811 & 0.0050 & C & \citep{2011IBVS.5977....1W} & - \\
55601.3466 & 0.0017 & -22377 & 0.0040 & C & \citep{2012IBVS.6015....1W} & - \\
56146.2889 & 0.0017 & -15971 & 0.0044 & C & \citep{2013IBVS.6049....1W} & - \\
56146.3731 & 0.0020 & -15970 & 0.0035 & C & \citep{2013IBVS.6049....1W} & - \\
56146.4587 & 0.0017 & -15969 & 0.0040 & C & \citep{2013IBVS.6049....1W} & - \\
56146.5427 & 0.0014 & -15968 & 0.0030 & C & \citep{2013IBVS.6049....1W} & - \\
56224.2954 & 0.0010 & -15054 & 0.0041 & C & \citep{2013IBVS.6049....1W} & - \\
58273.3938 & 0.0022 & 9034 & -0.0016 & CV & AAVSO & VMT \\
58303.5101 & 0.0014 & 9388 & 0.0009 & CV & AAVSO & VMT \\
58331.5812 & 0.0013 & 9718 & -0.0003 & CV & AAVSO & VMT \\
58377.6035 & 0.0010 & 10259 & 0.0005 & CV & AAVSO & VMT \\
58408.5678 & 0.0011 & 10623 & 0.0003 & CV & AAVSO & VMT \\
58436.3856 & 0.0010 & 10950 & 0.0010 & CV & AAVSO & VMT \\
58470.3286 & 0.0012 & 11349 & 0.0021 & CV & AAVSO & VMT \\
58504.5237 & 0.0024 & 11751 & 0.0002 & CV & AAVSO & VMT \\
58529.2796 & 0.0011 & 12042 & 0.0015 & CV & AAVSO & VMT \\
58564.4124 & 0.0017 & 12455 & 0.0014 & CV & AAVSO & VMT \\
59213.6486 & 0.0012 & 20087 & 0.0030 & CV & AAVSO & VMT \\
59214.6698 & 0.0018 & 20099 & 0.0034 & CV & AAVSO & VMT \\
59277.5335 & 0.0020 & 20838 & 0.0023 & CV & AAVSO & VMT \\
59414.4932 & 0.0010 & 22448 & 0.0035 & CV & AAVSO & HMB \\
59414.5789 & 0.0015 & 22449 & 0.0041 & CV & AAVSO & HMB \\
59439.5030 & 0.0013 & 22742 & 0.0034 & CV & AAVSO & HMB \\
59440.4398 & 0.0013 & 22753 & 0.0044 & CV & AAVSO & HMB \\
59440.5252 & 0.0009 & 22754 & 0.0048 & CV & AAVSO & HMB \\
59441.4597 & 0.0022 & 22765 & 0.0036 & CV & AAVSO & HMB \\
59443.4996 & 0.0019 & 22789 & 0.0018 & CV & AAVSO & HMB \\
59450.4773 & 0.0017 & 22871 & 0.0040 & CV & AAVSO & HMB \\
59451.4127 & 0.0023 & 22882 & 0.0037 & CV & AAVSO & HMB \\
59451.4964 & 0.0017 & 22883 & 0.0023 & CV & AAVSO & HMB \\
59457.4516 & 0.0022 & 22953 & 0.0028 & CV & AAVSO & HMB \\
59460.3471 & 0.0029 & 22987 & 0.0060 & CV & AAVSO & VMT \\
59460.4295 & 0.0011 & 22988 & 0.0034 & CV & AAVSO & HMB \\
59461.4503 & 0.0012 & 23000 & 0.0033 & CV & AAVSO & HMB \\
59462.3868 & 0.0024 & 23011 & 0.0040 & CV & AAVSO & HMB \\
59462.4716 & 0.0017 & 23012 & 0.0038 & CV & AAVSO & HMB \\
59463.4066 & 0.0019 & 23023 & 0.0031 & CV & AAVSO & HMB \\
59464.4303 & 0.0025 & 23035 & 0.0059 & CV & AAVSO & HMB \\
59464.5971 & 0.0006 & 23037 & 0.0026 & CV & AAVSO & VMT \\
59465.4490 & 0.0013 & 23047 & 0.0039 & CV & AAVSO & HMB \\
59465.6194 & 0.0012 & 23049 & 0.0042 & CV & AAVSO & VMT \\
59466.3828 & 0.0014 & 23058 & 0.0019 & CV & AAVSO & HMB \\
59466.4684 & 0.0019 & 23059 & 0.0024 & CV & AAVSO & HMB \\
59466.6394 & 0.0012 & 23061 & 0.0033 & CV & AAVSO & VMT \\
59479.3994 & 0.0027 & 23211 & 0.0032 & CV & AAVSO & HMB \\
59480.4208 & 0.0019 & 23223 & 0.0037 & CV & AAVSO & HMB \\
59496.3272 & 0.0017 & 23410 & 0.0026 & CV & AAVSO & HMB \\
59497.3480 & 0.0019 & 23422 & 0.0025 & CV & AAVSO & HMB \\
59500.3257 & 0.0020 & 23457 & 0.0029 & CV & AAVSO & HMB \\
59503.3876 & 0.0027 & 23493 & 0.0023 & CV & AAVSO & HMB \\
59507.3026 & 0.0015 & 23539 & 0.0043 & CV & AAVSO & HMB \\
59507.3870 & 0.0039 & 23540 & 0.0036 & CV & AAVSO & HMB \\
59511.2988 & 0.0022 & 23586 & 0.0024 & CV & AAVSO & HMB \\
59511.3843 & 0.0017 & 23587 & 0.0028 & CV & AAVSO & HMB \\
59514.1927 & 0.0020 & 23620 & 0.0039 & CV & AAVSO & PMAK \\
59514.2786 & 0.0015 & 23621 & 0.0047 & CV & AAVSO & PMAK \\
59515.2980 & 0.0017 & 23633 & 0.0034 & CV & AAVSO & HMB \\
59515.3826 & 0.0017 & 23634 & 0.0029 & CV & AAVSO & HMB \\
59516.2327 & 0.0019 & 23644 & 0.0023 & CV & AAVSO & HMB \\
59516.3189 & 0.0028 & 23645 & 0.0035 & CV & AAVSO & HMB \\
59516.4035 & 0.0013 & 23646 & 0.0030 & CV & AAVSO & HMB \\
59519.2098 & 0.0018 & 23679 & 0.0021 & CV & AAVSO & PMAK \\
59519.2951 & 0.0018 & 23680 & 0.0023 & CV & AAVSO & PMAK \\
59524.4024 & 0.0014 & 23740 & 0.0055 & CV & AAVSO & HMB \\
59527.2930 & 0.0015 & 23774 & 0.0038 & CV & AAVSO & HMB \\
59527.3776 & 0.0022 & 23775 & 0.0034 & CV & AAVSO & HMB \\
59528.2289 & 0.0011 & 23785 & 0.0040 & CV & AAVSO & HMB \\
59528.3136 & 0.0015 & 23786 & 0.0036 & CV & AAVSO & HMB \\
59528.3988 & 0.0010 & 23787 & 0.0037 & CV & AAVSO & HMB \\
59529.2483 & 0.0013 & 23797 & 0.0026 & CV & AAVSO & HMB \\
59529.3333 & 0.0013 & 23798 & 0.0025 & CV & AAVSO & HMB \\
59529.4186 & 0.0009 & 23799 & 0.0028 & CV & AAVSO & HMB \\
59533.4177 & 0.0011 & 23846 & 0.0037 & CV & AAVSO & HMB \\
59536.2256 & 0.0021 & 23879 & 0.0043 & CV & AAVSO & HMB \\
59536.3091 & 0.0050 & 23880 & 0.0028 & CV & AAVSO & HMB \\
59540.2237 & 0.0014 & 23926 & 0.0043 & CV & AAVSO & HMB \\
59540.3092 & 0.0015 & 23927 & 0.0047 & CV & AAVSO & HMB \\
59540.3944 & 0.0010 & 23928 & 0.0048 & CV & AAVSO & HMB \\
59540.4779 & 0.0006 & 23929 & 0.0033 & CV & AAVSO & HMB \\
59551.2801 & 0.0012 & 24056 & 0.0019 & CV & AAVSO & HMB \\
59557.2367 & 0.0009 & 24126 & 0.0038 & CV & AAVSO & HMB \\
59557.3212 & 0.0010 & 24127 & 0.0032 & CV & AAVSO & HMB \\
59557.4074 & 0.0011 & 24128 & 0.0044 & CV & AAVSO & HMB \\
59557.4916 & 0.0011 & 24129 & 0.0035 & CV & AAVSO & HMB \\
59565.2334 & 0.0014 & 24220 & 0.0042 & CV & AAVSO & HMB \\
59565.3182 & 0.0013 & 24221 & 0.0039 & CV & AAVSO & HMB \\
59565.4029 & 0.0009 & 24222 & 0.0035 & CV & AAVSO & HMB \\
59569.2312 & 0.0016 & 24267 & 0.0038 & CV & AAVSO & HMB \\
59569.3157 & 0.0014 & 24268 & 0.0032 & CV & AAVSO & HMB \\
59569.4011 & 0.0012 & 24269 & 0.0035 & CV & AAVSO & HMB \\
59569.4858 & 0.0009 & 24270 & 0.0032 & CV & AAVSO & HMB \\
59589.2216 & 0.0015 & 24502 & 0.0033 & CV & AAVSO & HMB \\
59589.3050 & 0.0009 & 24503 & 0.0017 & CV & AAVSO & HMB \\
59591.3484 & 0.0021 & 24527 & 0.0035 & CV & AAVSO & HMB \\
59598.3232 & 0.0005 & 24609 & 0.0027 & CV & AAVSO & HMB \\
59604.2781 & 0.0017 & 24679 & 0.0029 & CV & AAVSO & HMB \\
59604.3638 & 0.0014 & 24680 & 0.0035 & CV & AAVSO & HMB \\
59622.3132 & 0.0013 & 24891 & 0.0037 & CV & AAVSO & HMB \\
59623.3331 & 0.0012 & 24903 & 0.0028 & CV & AAVSO & HMB \\
59625.2914 & 0.0021 & 24926 & 0.0046 & CV & AAVSO & HMB \\
59636.2636 & 0.0012 & 25055 & 0.0031 & CV & AAVSO & HMB \\
59637.2836 & 0.0016 & 25067 & 0.0023 & CV & AAVSO & HMB \\
59638.3057 & 0.0019 & 25079 & 0.0035 & CV & AAVSO & HMB \\
59639.3255 & 0.0011 & 25091 & 0.0025 & CV & AAVSO & HMB \\
59641.2828 & 0.0015 & 25114 & 0.0032 & CV & AAVSO & HMB \\
59642.3026 & 0.0011 & 25126 & 0.0022 & CV & AAVSO & HMB \\
59643.3243 & 0.0020 & 25138 & 0.0032 & CV & AAVSO & HMB \\
59644.3453 & 0.0014 & 25150 & 0.0033 & CV & AAVSO & HMB \\
59646.3021 & 0.0011 & 25173 & 0.0036 & CV & AAVSO & HMB \\
59647.3237 & 0.0012 & 25185 & 0.0044 & CV & AAVSO & HMB \\
59648.3433 & 0.0010 & 25197 & 0.0032 & CV & AAVSO & HMB \\
59649.2791 & 0.0015 & 25208 & 0.0033 & CV & AAVSO & HMB \\
59658.2960 & 0.0012 & 25314 & 0.0030 & CV & AAVSO & HMB \\
59659.3162 & 0.0015 & 25326 & 0.0024 & CV & AAVSO & HMB \\
59660.3374 & 0.0014 & 25338 & 0.0028 & CV & AAVSO & HMB \\
59662.2945 & 0.0017 & 25361 & 0.0033 & CV & AAVSO & HMB \\
59663.3158 & 0.0012 & 25373 & 0.0038 & CV & AAVSO & HMB \\
59664.3361 & 0.0012 & 25385 & 0.0034 & CV & AAVSO & HMB \\
59783.3462 & 0.0012 & 26784 & 0.0040 & CV & AAVSO & PMAK \\
59810.3115 & 0.0007 & 27101 & 0.0031 & CV & AAVSO & PMAK \\
59854.2910 & 0.0009 & 27618 & 0.0027 & CV & AAVSO & HMB \\
59854.3749 & 0.0010 & 27619 & 0.0015 & CV & AAVSO & HMB \\
59860.2456 & 0.0012 & 27688 & 0.0025 & CV & AAVSO & PMAK \\
59860.3314 & 0.0015 & 27689 & 0.0032 & CV & AAVSO & PMAK \\
59933.2350 & 0.0010 & 28546 & 0.0041 & CV & AAVSO & PMAK \\
59933.3191 & 0.0008 & 28547 & 0.0032 & CV & AAVSO & PMAK \\
59981.4681 & 0.0020 & 29113 & 0.0040 & CV & AAVSO & SDOA \\
60005.4556 & 0.0013 & 29395 & 0.0024 & CV & AAVSO & SDOA \\
60038.4621 & 0.0010 & 29783 & 0.0028 & CV & AAVSO & SDOA \\
60170.3175 & 0.0012 & 31333 & 0.0036 & CV & AAVSO & PMAK \\
60170.4030 & 0.0009 & 31334 & 0.0041 & CV & AAVSO & PMAK \\
60172.3583 & 0.0010 & 31357 & 0.0028 & CV & AAVSO & PMAK \\
60192.6039 & 0.0009 & 31595 & 0.0024 & CV & AAVSO & SDOA \\
60197.2830 & 0.0010 & 31650 & 0.0028 & CV & AAVSO & PMAK \\
60197.3687 & 0.0010 & 31651 & 0.0035 & CV & AAVSO & PMAK \\
60219.5693 & 0.0015 & 31912 & 0.0014 & CV & AAVSO & SDOA \\
60235.2223 & 0.0015 & 32096 & 0.0020 & CV & AAVSO & PMAK \\
60235.3077 & 0.0015 & 32097 & 0.0024 & CV & AAVSO & PMAK \\
60320.3757 & 0.0009 & 33097 & 0.0030 & CV & AAVSO & SDOA \\
60377.4558 & 0.0010 & 33768 & 0.0028 & CV & AAVSO & SDOA \\
60378.3064 & 0.0009 & 33778 & 0.0027 & CV & AAVSO & PMAK \\
60378.3920 & 0.0017 & 33779 & 0.0033 & CV & AAVSO & PMAK \\
60410.2916 & 0.0011 & 34154 & 0.0026 & CV & AAVSO & PMAK \\
60456.3981 & 0.0011 & 34696 & 0.0026 & CV & AAVSO & PMAK \\
60480.3867 & 0.0014 & 34978 & 0.0022 & CV & AAVSO & PMAK \\
\hline
\label{tab1}
\end{longtable}
}

\end{document}